\documentclass[aps,prd,twocolumn,nofootinbib,superscriptaddress]{revtex4-2}

\usepackage{amsbsy}
\usepackage{multirow}
\usepackage{amssymb}
\usepackage{amsmath}
\usepackage{graphicx}
\usepackage{color}
\usepackage{verbatim}

\setcitestyle{square}
\usepackage{caption}
\usepackage{hyperref,url}

\usepackage{subcaption}
\usepackage{mwe}

\begin{document}

\title{Beijing Normal University 12-meter Interferometric kHz GW Detector Prototype: Design and Scientific Prospects}

\author{Mengyao Wang}
\email{mengyao.wang@bnu.edu.cn}
\affiliation{School of Physics and Astronomy, Beijing Normal University, Beijing, 100875, China}

\author{Fan Zhang}
\email{fnzhang@bnu.edu.cn}
\affiliation{Institute for Frontiers in Astronomy and Astrophysics, Beijing Normal University, Beijing 102206, China}
\affiliation{School of Physics and Astronomy, Beijing Normal University, Beijing, 100875, China}
\affiliation{Advanced Institute of Natural Sciences, Beijing Normal University at Zhuhai, Zhuhai, 519087, China}
\author{Xinyao Guo}
\author{Haixing Miao}
\affiliation{Frontier Science Center for Quantum Information, Department of Physics, Tsinghua University, Beijing, 100084, China}
\author{Huan Yang}
\affiliation{Department of Astronomy, Tsinghua University, Beijing, 100084, China}
\author{Yiqiu Ma}
\affiliation{National Gravitation Laboratory, MOE Key Laboratory of Fundamental Physical Quantities Measurement, School of Physics, Huazhong University of Science and Techonology, Wuhan, 430074, China}
\author{Haoyu Wang}
\affiliation{Department of Physics, Institute of Science Tokyo, 2-12-1 Ookayama, Meguro-ku, Tokyo 152-8551, Japan}
\author{Haibo Wang}
\affiliation{School of Physics and Astronomy, Beijing Normal University, Beijing, 100875, China}
\author{Teng Zhang}
\affiliation{School of Physcis and Astronomy, University of Birmingham, Birmingham, UK, B15 2TT}
\author{Mengdi Cao}
\author{Yuchao Chen}
\author{Xiaoman Huang}
\author{Fangfei Liu}
\author{Jianyu Liu}
\author{Yuan Pan}
\affiliation{School of Physics and Astronomy, Beijing Normal University, Beijing, 100875, China}
\author{Junlang Li}
\affiliation{School of Physics and Technology, Wuhan University, Wuhan 430072, China}
\affiliation{School of Physics and Astronomy, Beijing Normal University, Beijing, 100875, China}
\author{Yulin Xia}
\affiliation{Frontier Science Center for Quantum Information, Department of Physics, Tsinghua University, Beijing, 100084, China}
\author{Jianbo Xing}
\author{Yujie Yu}
\author{Chenjie Zhou}
\affiliation{School of Physics and Astronomy, Beijing Normal University, Beijing, 100875, China}
\author{Zong-Hong Zhu}
\email{zhuzh@bnu.edu.cn}
\affiliation{Institute for Frontiers in Astronomy and Astrophysics, Beijing Normal University, Beijing 102206, China}
\affiliation{School of Physics and Astronomy, Beijing Normal University, Beijing, 100875, China}

%\ead{mengyao.wang@bnu.edu.cn, fnzhang@bnu.edu.cn}

\begin{abstract}
Current gravitational-wave detectors have achieved remarkable sensitivity around 100 Hz, enabling ground-breaking discoveries. Enhancing sensitivity at higher frequencies in the kilohertz (kHz) range promises access to rich physics, particularly the extreme conditions during the merger stage of binary neutron stars. However, the high-frequency sensitivity of Michelson-based interferometers is fundamentally limited by their linear optical cavities, which are optimized for low-frequency signal enhancement. In [Phys. Rev. X 13, 021019 (2023)], a new configuration employing an L-shaped optical resonator was proposed to overcome this limitation, offering exceptional sensitivity in the kHz band.
As a pathfinder, the 12-meter prototype at Beijing Normal University is designed to demonstrate the sensing and control schemes of this new kHz detector configuration and to explore its performance in the high-power regime with suspended optics. Beyond its primary scientific goal, the prototype also offers potential sensitivity in the megahertz (MHz) range, potentially enabling constraints on exotic sources.
This paper presents an overview of the prototype, including its optical design and current development status of key components. 
\end{abstract}

\maketitle

\section{Introduction}

The emerging field of gravitational wave (GW) astronomy holds great promise, offering insights into processes occurring in regions of extreme energy density\,\cite{2016PhRvL.116f1102A,PhysRevLett.119.161101,AdLigo2015}, thus providing information complementary to electromagnetic observations. The GW170817 event from a binary neutron star merger exemplifies how multi-messenger observations can lead to groundbreaking discoveries\,\cite{Abbott_2017,Abbott_HC,Martynov19, Bailes21}.

A central objective in the future of gravitational-wave (GW) astronomy is to expand the detection frequency band to encompass both the low- and high-frequency extremes of the GW spectrum\,\cite{CMB, AmaroSeoane2023, tianqin, Taiji, Li2023}. For ground-based detectors, improving sensitivity in the kilohertz (kHz) regime is a key part of this effort. One of the primary motivations arises from the GW170817 event, in which the merger phase of the signal lies in the kHz range—beyond the most sensitive band of current detectors such as LIGO, Virgo, and KAGRA. Yet this merger signal encodes crucial information about the internal structure and equation of state of neutron stars. Achieving high sensitivity in the kHz regime is therefore of great scientific importance.

Several proposals aim to improve kHz sensitivity by modifying the input/output optics of the Michelson interferometer configuration. However, the linear optical cavities in the Michelson interferometer are inherently optimized to enhance signals at frequencies below the cavity bandwidth. Consequently, optical losses occurring after the arm cavity, especially those associated with the signal-recycling cavity, severely limit high-frequency sensitivity by introducing vacuum fluctuations that degrade quantum coherence. Importantly, this quantum loss limitation cannot be effectively mitigated by increasing the arm length.

In Ref.\,\cite{zhang2023}, a new kHz detector configuration employing an L-shaped optical resonator as the arm cavity has been proposed to overcome this limitation. This configuration allows for direct amplification of kHz signals within the arm cavities, significantly enhancing sensitivity to binary neutron star merger signals at these frequencies. Extensive research and development are required to validate and refine this approach, ensuring its viability as a future kHz GW detector. The first step involves testing the sensing and control schemes, as well as evaluating high-power operation with suspended optics.

Complementary to other prototypes in the community\,\cite{Caltech40m, Goßler_2010, Zhao_2006, Glasgow10}, The Beijing Normal University (BNU) prototype is dedicated specifically to addressing technical challenges associated with this new detector configuration, e.g., dark-port injection and the implications of leaving one degree of freedom uncontrolled. Its primary aim is to demonstrate the feasibility of lock acquisition and the sensing-control scheme using suspended optics within a 12-meter scale L-shaped vacuum envelope. Additionally, once operational, the prototype will enable investigation into challenges related to high-power operation, laser-noise coupling, and squeezed light injection,  critical elements for kHz detection. 

Beyond its principal role as a testbed for kHz detector technology, the BNU prototype is also expected to exhibit potential sensitivity in the MHz range. Exploring the GW background in this frequency domain could reveal exotic astrophysical sources or signals originating from unknown high-energy physics phenomena. As such, the prototype has the potential to provide valuable constraints complementary to other detection techniques\,\cite{Aggarwal:2020olq}.

%Prototypes like the Caltech 40m\cite{Caltech40m}, AEI 10m\cite{ Goßler_2010}, Gingin 80\,m\cite{Zhao_2006} and Glasgow 10m\cite{Glasgow10}, enable scientists and engineers to gain expertise, refine techniques, and ensure a seamless transition to operational large-scale detectors. Such efforts also promote international collaboration and serve as training platforms for aspiring GW scientists and engineers.

This paper is to provide an overview of the BNU prototype, detailing its design and key components.  It is organized as follows: Section\,\ref{secII} reviews key aspects of the kHz detector configuration utilizing an L-shaped resonator, outlining the overall scientific goals of the prototype. Section\,\ref{secIII} presents the optical design and key parameters of the prototype. Section\,\ref{secIV} describes additional critical components, including the vacuum system, and seismic isolation, and digital control systems. Section\,\ref{secV} shows its MHz sensitivity and offers a preliminary exploration of the implications. 
Finally, we conclude and provide an outlook in Section\,\ref{sect:conclusion}. 

%% Section II goal

\section{Review of the kHz detector design with L-shaped resonator}
\label{secII}

\begin{figure}[!b]
\includegraphics[width=0.48\textwidth]{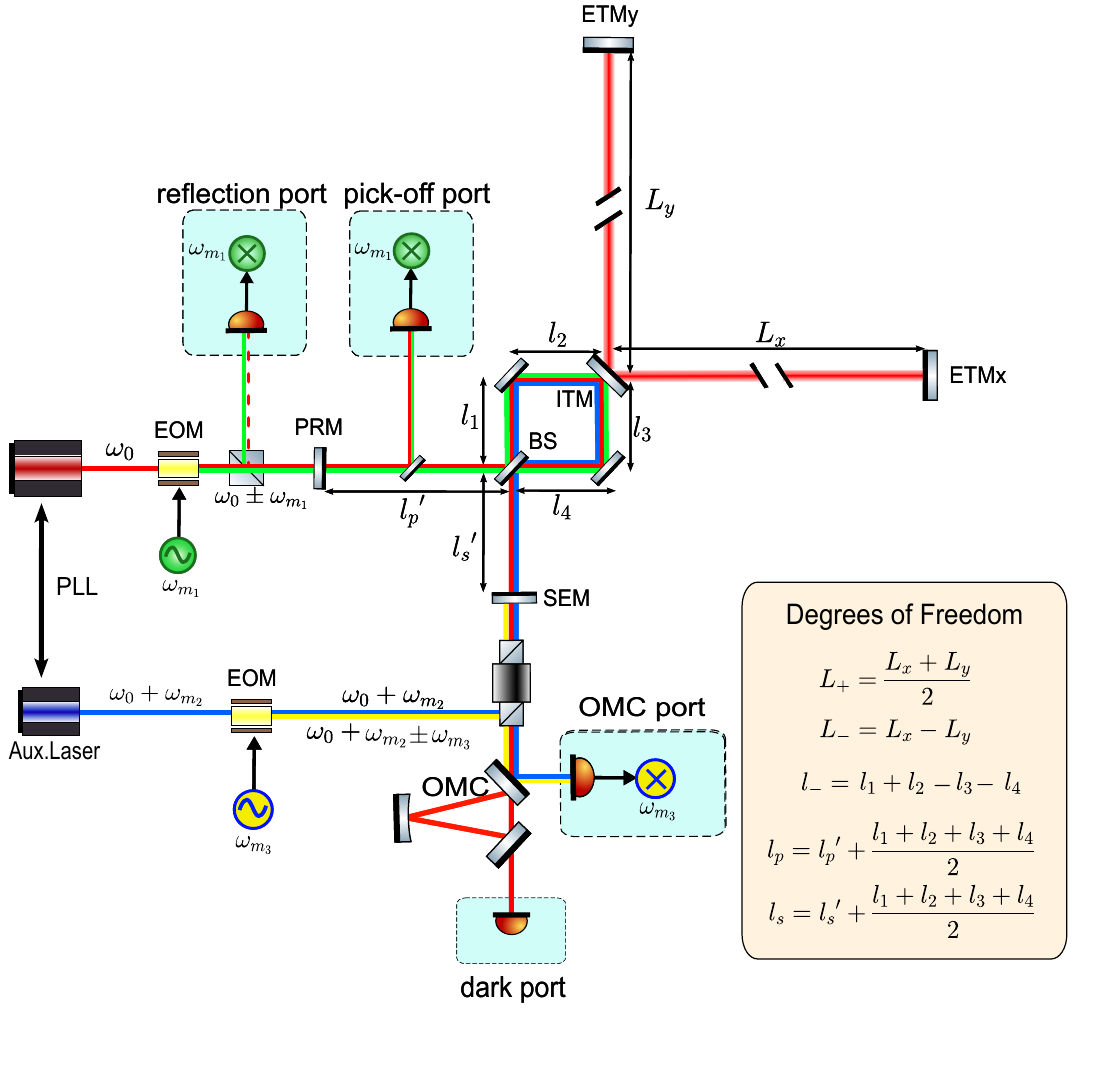}
        \caption[r]
{Illustration of the new kHz detector configuration and its sensing and control scheme. The corresponding demodulation frequencies and readout ports are also highlighted. EOM: electro-optic modulator; PRM: power recycling mirror; ITM: input test mass; BS: beam splitter; ETM: end test mass; OMC: output mode cleaner; PLL: phase-locked loop.}
        \label{fig:scheme}
\end{figure}
The core of the new kHz detector design consists of an L-shaped optical resonator acting as the arm cavity, enabling direct amplification of the GW signal near its first free spectral range. This design allows for overcoming the quantum loss limit, which significantly constrains conventional Michelson configurations. Consequently, a notable sensitivity enhancement at frequencies in the kHz regime can be achieved.

An intuitive, though simplified, explanation of this configuration's frequency characteristics is as follows. Photons in the resonator can propagate in two orthogonal directions, labeled $x$ and $y$ (see Fig.~\ref{fig:scheme} for optical layout details). Consider one photon: it first traverses the $x$ direction followed by the $y$ direction. If a GW alternates at a frequency around $c/L$ (where $L$ is the arm length scale), then as it travels along the $x$ ($y$) direction, this direction is either stretched or squeezed. Subsequently, when the photon moves into the perpendicular direction ($y$ or $x$), the GW-induced deformation has accumulated an additional $\pi$ phase shift, causing that direction also to be correspondingly stretched or squeezed. Thus, the accumulated phase coherently adds up for every round-trip, leading to a signal amplification. For a full-scale detector with tens-of-kilometers-long arms, this optimal GW frequency conveniently lies within the scientifically compelling kHz frequency range.

A significant challenge in implementing the L-shaped resonator configuration involves achieving a stable lock of the complete optical setup. Due to the fundamental differences between the optical layouts of the new configuration and the traditional Michelson interferometer, the overall locking strategy requires a comprehensive redesign. This effort encompasses two critical aspects: linear sensing and control around the operating point, and lock acquisition—the procedure of transitioning the system from an initially free-swinging state to the proximity of the operating point, at which point linear control methods become effective.

For linear locking around the operating point, a preliminary control scheme has been developed\,\cite{Guo2023}, as shown in Fig.\,\ref{fig:scheme}. This scheme differs fundamentally from the conventional Michelson control approach in two key aspects. 
First, the control field is injected through the dark port. Due to the system’s Sagnac-like optical response for the control sidebands, cross-coupling between the bright and dark ports is fully suppressed. As a result, a pair of auxiliary fields must be injected through the dark port to monitor the length fluctuations of the signal extraction cavity.
Second, the system has one fewer degree of freedom requiring active control. At low frequencies, two differential modes—denoted 
$L_-$ and $l_-$—become fully degenerate. Only one linear combination, namely 
$L_-+l_-$ , produces a non-zero optical response and thus requires control. The orthogonal combination is completely decoupled from the light field and remains hidden. Consequently, although the system nominally has five longitudinal degrees of freedom, only four require active feedback control, with the hidden mode left free-swinging.

However, this control design remains conceptual. The feasibility of dark-port injection and the implications of leaving one degree of freedom uncontrolled have yet to be experimentally demonstrated—neither in tabletop experiments nor at the prototype scale. As an experimental platform, the BNU prototype is ideally positioned to validate the viability of the lock acquisition and this new sensing and control scheme, thereby paving the way for the future construction of a full-scale  GW detector with L-shaped resonator.

%% Section III design

\section{Optical design and key parameters}
\label{secIII}

 \begin{figure}[!b]
 \centering
\includegraphics[width=0.99\columnwidth]{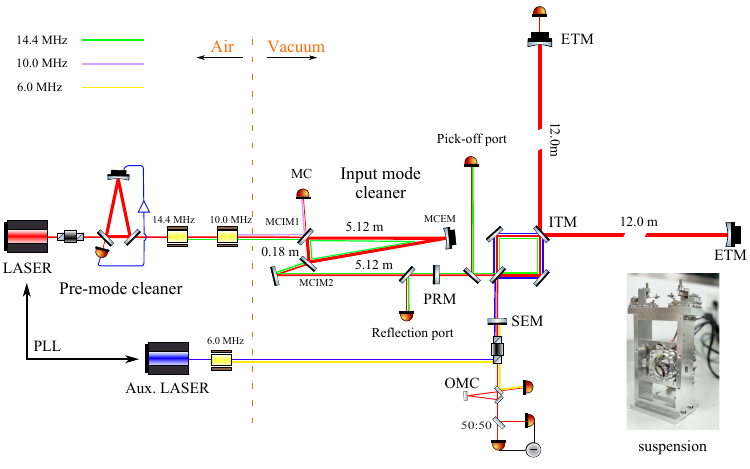}
    \caption{Optical layout of the 12-meter BNU prototype. The core optics are suspended using LIGO-type tip-tilt suspensions (see inset)\cite{tiptilt}.}
    \label{fig:L-shape}
\end{figure}

The optical design of the BNU prototype is shown in Fig.\,\ref{fig:L-shape}, and key parameters are summarized in Table\,\ref{table:paras}. Its core components are arranged following the layout of the new kHz detector configuration illustrated in Fig.\,\ref{fig:scheme}—an L-shaped resonator combined with a Sagnac-type ring loop. The experimental demonstrations are planned to be conducted in two distinct stages, each corresponding to a slightly different optical configuration. This staged approach is crucial for systematically testing and validating key technologies.

\begin{table*}[!t]
\centering
\vspace{0.8cm}
\begin{tabular}{|p{4cm}|p{7cm}|p{4cm}|}
\hline
    \textbf{Item}&  \textbf{Parameters}&  \textbf{Value}\\

\hline
    Laser &Input power & 15\,W\\
          & Wavelength & 1064\,nm\\
    EOM & Modulation frequency 1 (cavity locking)& 14.4\,MHz \\
        & Modulation frequency 2 (IMC locking) & 10\,MHz\\
\hline
    Input Mode Cleaner (IMC) & Length & 10\,m\\
                       & MCEM loss & 50\,ppm\\
                       & Transmissivity of MCIM1 and MCIM2 & 0.001 (s-polarization)\\
                       & & 0.02 (p-polarization)\\ 
\hline
    L-shaped Cavity & Arm length & 12\,m\\ 
            & ETMs and ITM diameter & 2\,inch\\
            & ETMs RoC & 17.0\,m\\
            & ITM RoC & $\infty$\\
            & ETM loss & 50\,ppm\\
            & ITM transmissivity & 0.02\\
            & Circulating power & 300\,kW\\
\hline
    PRM & Transmissivity  & 0.01\\
        & Path length to ITM & 5.2\,m\\
        & RoC & 16.7\,m\\   
        & Circulating power & 1500\,W\\
\hline
    SEM & Transmissivity & 0.1\\
        & Path length to ITM & 5.4\,m\\
        & RoC & 16.5\,m\\
        & Auxiliary laser shift frequency & 27.7\,MHz\\
        & EOM modulation frequency& 6.0\,MHz\\ 
\hline
\end{tabular}
\caption{Key parameters of the BNU prototype.}
\label{table:paras}
\end{table*}

\textbf{First Stage — Arm Cavity Locking:} The initial stage focuses on locking a suspended L-shaped arm cavity. This cavity is formed by a plano ITM with 45-degree incidence and ETMs, with transmissivities of 0.01 and 50\,ppm, respectively; the corresponding cavity finesse is approximately 300. 

\begin{figure}[!b]
    \centering   \includegraphics[width=0.9\columnwidth]{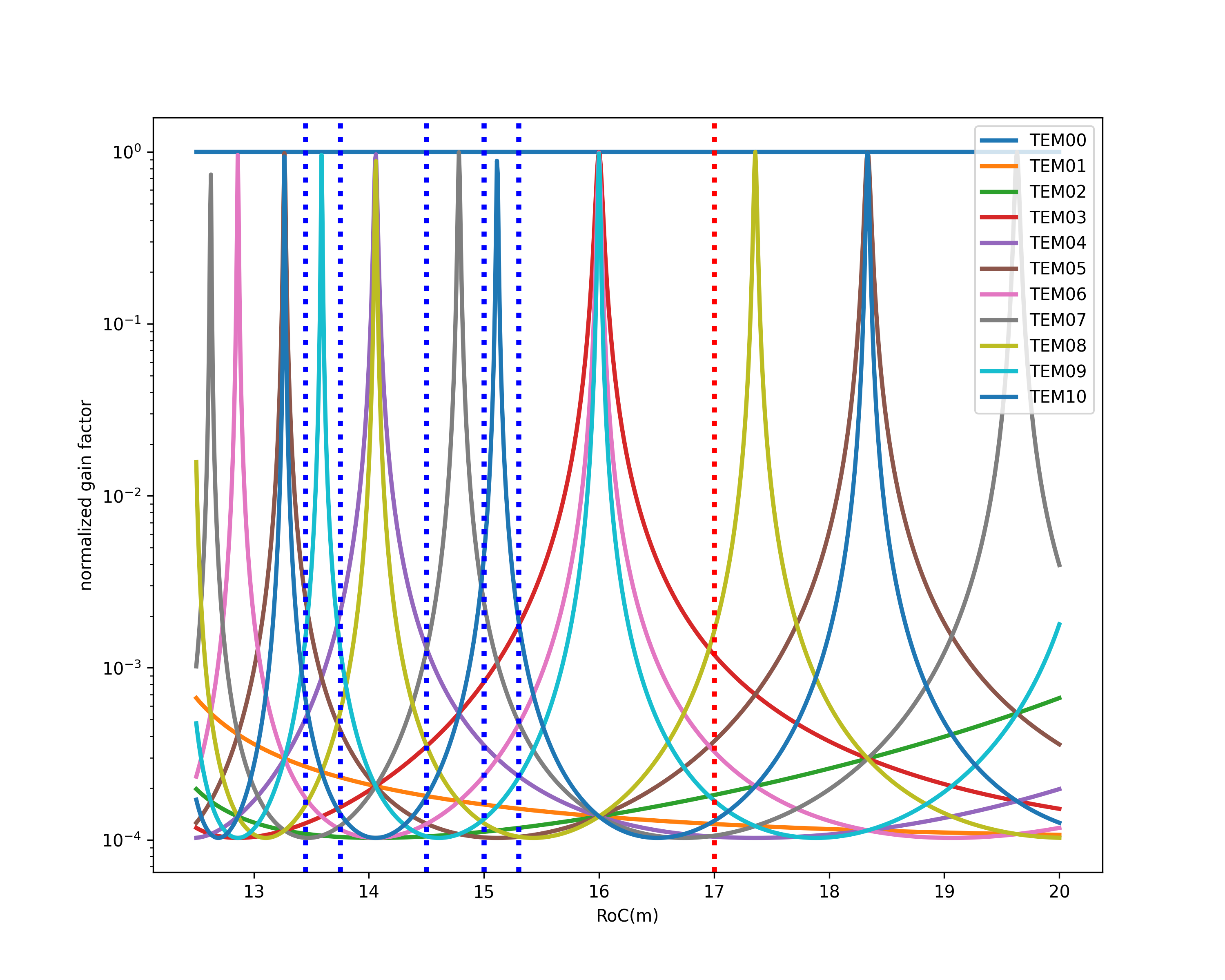}
    \caption{Normalized gain factor (normalized by the maximum gain of the fundamental mode) of different higher-order modes, as a function of the ETM RoC.}
    \label{fig:roc_scan}
\end{figure}

To suppress higher-order mode coupling, we performed a scan of the ETM radius of curvature, as illustrated in Fig.\,\ref{fig:roc_scan} where the optical gain for various higher-order modes are plotted.
Among the available options, a curvature radius of \(17.0\,\mathrm{m}\) was selected, as it provides effective suppression of the higher-order modes. The resulting g-factor of the cavity is $ -0.29$, and the one-way Guoy phase is approximately equal to 1.3 radian. 

In this stage, the main scientific objectives are two-fold. First, the non-adiabatic effect associated with the hidden degree of freedom---arising from the degeneracy of $l_-$ and $L_-$ mentioned earlier---is expected to manifest in this layout. This will  enable experimental validation of the strategy to leave one degree of freedom free-swinging. Second, the simplified optical layout offers critical insights into the coupling dynamics between Michelson- and Sagnac-like modes far from the operating point, thereby informing the design and implementation of full-degree-of-freedom lock acquisition protocols.

%\textcolor{black}{Xinyao: Notably, imbalance of two folds of L-shaped arm cavity would not affect the sensing control of system at low frequency. Given the spatial constraints, we adopted an asymmetric L-shaped arm with horizontal fold of 12m and vertical fold of 10 m in the preliminary locking stage.} 

\textbf{Second Stage — Recycling Cavity Formation and Control:} Upon successful completion of the first stage, the experiment will proceed to the second phase, which involves introducing additional  mirrors—the power recycling mirror (PRM) and the signal extraction mirror (SEM). This enhances the circulating power and tunes the overall bandwidth. The primary objective of this stage is to implement and evaluate the full sensing and control scheme proposed in Ref.\,\cite{Guo2023}, and and to investigate dynamical effects arising in the high-power operational regime.

\begin{table*}[t!]
\begin{center}
\vspace{0cm}
\begin{tabular}{cccc}
\hline
Transfer Functions& Expression& value&unit\\
\hline
${\cal T}_{\,\,L_+}^{\rm refl}$& $
\frac{ \pi  t_{\rm i}^2\rm P_0 \beta_1}{\lambda_0 (1-r_{\rm i}r_{\rm e})^2} (g_{\rm p}^{\rm c})^2r_{\rm p}^{\rm s}\times\frac{1}{1+if/f_{\rm cc}}$&$\frac{7.7\times10^{10}}{1+(f/156{\rm Hz})i}$&$\rm W/m$\\

${\cal T}_{\,\,L_+}^{\rm pop}$& $
\frac{ \pi t_{\mathrm{po}}^2 t_{\rm i}^2\rm P_0 \beta_1}{\lambda_0 t_p(1-r_{\rm i}r_{\rm e})^2} (g_{\rm p}^{\rm c})^2g_{\rm p}^{\rm s}\times\frac{1}{1+if/f_{\rm cc}}$&$\frac{4.1\times10^9}{1+(f/156{\rm Hz})i}$&$\rm W/m$\\

${\cal T}_{\,\,l_p}^{\rm refl}$&$\frac{2\pi \rm P_0 \beta_1}{\lambda_0}\left[(g_{\rm p}^{\rm c})^2\,r_{\rm p}^{\rm s}-(g_{\rm p}^{\rm s})^2\,r_{\rm p}^{\rm c} \right]\times \frac{1+if/f_{\rm r}}{1+if/f_{\rm cc}}$&$7.2\times10^8\times\frac{1+(f/3569{\rm Hz})i}{1+(f/156{\rm Hz})i}$&$\rm W/m$\\

${\cal T}_{\,\,l_p}^{\rm pop}$&$\frac{2\pi \rm P_0 \beta_1t_{\rm po}^2}{\lambda_0 t_p}g_{\rm p}^{\rm c}\,g_{\rm p}^{\rm s}(g_{\rm p}^{\rm c}-g_{\rm p}^{\rm s}) \times \frac{1+if/f_{\rm p}}{1+if/f_{\rm cc}}$&$3.8\times10^7\times\frac{1+(f/820{\rm Hz})i}{1+(f/156{\rm Hz})i}$&$\rm W/m$ \\

${\cal T}_{\,\,l_s}^{\rm omc}$&$\frac{2\pi {\rm P_a}\beta_2}{\lambda_0}\times r_s\, g_{\rm s}^{\rm s}$&$3.6\times10^6$&$\rm W/m$  \\

${\cal T}_{\,\,\cal L_-}^{\rm dc}$&$\frac{4\pi \rm P_0 \sin(\Delta \phi)}{\lambda_0}\times \frac{(t_{\rm i}\, g_{\rm p}^{\rm c}\, g_{\rm s}^{\rm c})^2}{(1+r_{\rm i}r_{\rm e})^2}$&$1.7\times10^9$&$\rm W/m$  \\
\hline
Important Poles\\
$f_{\rm cc}$& $\frac{c}{16 L_{+}}(1-r_i r_e)(1-r_p r_{\mathrm{L}, \rm c}^{(+)})$&157&$\rm Hz$  \\
$f_{\rm r}$& $f_{\rm c c}\left[1-\frac{(g_{\rm p}^{\rm c})^2\, r_{\mathrm{p}}^ \mathrm{s}}{(g_{\rm p}^{\rm s})^2\, r_{\mathrm{p}}^ \mathrm{c}}\right]$&3569&$\rm Hz$ \\

$f_{\rm p}$& $f_{\rm c c}\left(\frac{g_{\rm p}^{\rm c}}{g_{\rm p}^{\rm s}}-1\right)$&820&$\rm Hz$ \\
\hline

% Characteristic Gain & Definition & Effective reflectivity & Definition \\  
% $g_\textrm{p}^{\rm c}$& $\frac{t_{\rm p}}{1+t_{\rm po}^2r_{\rm p}r_\textrm{L,\,c}^{(+)}}$& & \\
% $g_\textrm{p}^{\rm s}$& $\frac{r_{\rm p}}{1-t_{\rm po}^2r_{\rm p}r_\textrm{L,\,s}^{(+)}}$& & \\ 
% $g_\textrm{s}^{\rm c}$& $\frac{t_{\rm s}}{1+r_{\rm s}r_\textrm{L,\,c}^{(-)}}$& &\\ $g_\textrm{s}^{\rm s}$&$\frac{t_{\rm s}}{1+r_{\rm s}r_\textrm{L,\,s}^{(-)}}$&&\\
\hline
\end{tabular}
\caption{Analytical expressions and numerical values of non-zero transfer functions in the sensing matrix. Here, $\rm P_0$  and $\rm P_{a}$  denote the powers of the main and auxiliary lasers, respectively, and $\beta_1 \,,\beta_2$ are the corresponding modulation depth of sidebands. $\Delta \cal L_-$ represents the DC offset of differential degree of freedom, and $\Delta \phi$ is the resulting excess phase of the carrier at the dark port. Detailed definitions of the symbols used in the table are provided in the appendix.}\label{tab:TFs}
\end{center}
\end{table*}

\newpage
\newpage
Key design considerations are as follows:
\begin{itemize}
  \item[(a)] \textbf{Critical coupling of the power-recycled arm cavity}
  
  The power-recycling mirror (PRM) reflectivity \(r_{\mathrm p}\) and the arm-cavity finesse are chosen to satisfy the critical coupling condition:
  \begin{equation}
    r_{\mathrm p}=r_{\rm eff}=\frac{r_e-r_i}{1-r_i r_e}, 
  \end{equation}
  where $
r_{\rm eff}$
  is the effective reflectivity of the L-shaped arm, and $r_i$ and $r_e$ are the amplitude relectivity of ITM and ETM, repsectively. The arm circulating power is related to the input laser power by:
\begin{equation}
    P_{\rm arm}=\left[\frac{1}{1-r_e^2}-\left(\frac{r_{\rm eff}-r_{\rm p}}{1-r_{\rm p}r_{\rm eff}}\right)^2\right]P_{\rm in}\,.
\end{equation}
    Critical coupling maximizes the circulating power in the arms for a fixed input laser power, and also results in a cleaner control signal at the reflection port. 
    
  \item[(b)] \textbf{Modulation frequency of bright-port sideband}

  The bright-port modulation frequency is set to half the free spectral range of the power-recycling cavity,
  \[
    \frac{\omega_{m_1}}{2\pi}=14.4\,\mathrm{MHz}=\frac{c}{4l_p},
  \]
  ensuring that the modulation sideband is resonant in the PRC. The PRC length is chosen as
  \(l_p=5.2\,\mathrm{m}\), placing the modulation frequency safely away from the
  arm cavity FSRs at \(12.5\,\mathrm{MHz}\) and \(18.75\,\mathrm{MHz}\), as required in Ref.\,\cite{Guo2023}. This sideband is used to sense both the PRC length $l_p$ and the common arm length $L_+$ in the control scheme.

  \item[(c)] \textbf{Modulation frequency of auxiliary laser and dark-port sideband}
  
  Similarly, the choice of auxiliary laser frequency shift and signal extraction cavity (SEC) length
  \[
    \frac{\omega_{m_2}}{2\pi}=27.7\,\mathrm{MHz}=\frac{c}{2l_s},
    \qquad l_s=5.4\,\mathrm{m},
  \]
  ensures that the auxiliary laser is resonant exclusively in the SEC. The modulation frequency at the dark port is selected as 
  \[
    \frac{\omega_{m_3}}{2\pi}=6\,\mathrm{MHz}
  \]
  so that the sidbands at 
  \(\bigl(\omega_{m_2}\!\pm\!\omega_{m_3}\bigr)/2\pi\) are off-resonant in all
  cavities within the system.  Demodulating the dark-port power at \(\omega_{m_3}\) then yields a error signal for controlling the SEC length \(l_s\).
\end{itemize}

We further validated the experimental feasibility of testing the control scheme through quantitative frequency-domain transfer function analysis. The relationship between length fluctuations and power readouts at each detection port is characterized by:
\begin{equation}
\left(\begin{array}{c}
\tilde{P}_{\text {refl }} \\
\tilde{P}_{\text {pop }} \\
\tilde{P}_{\text {omc }} \\
\tilde{P}_{\text {dc }}
\end{array}\right)=\left(\begin{array}{cccc}
{\cal T}_{\,\,L_+}^{\rm refl}  & {\cal T}_{\,\,l_p}^{\rm refl}  & 0 & 0 \\
{\cal T}_{\,\,L_+}^{\rm pop} & {\cal T}_{\,\,l_p}^{\rm pop} & 0 & 0 \\
0 & 0 & {\cal T}_{\,\,l_s}^{\rm omc} & 0 \\
0 & 0 & 0 & {\cal T}_{\,\,\cal L_-}^{\rm dc}
\end{array}\right) \quad\left(\begin{array}{c}
L_{+} \\
l_p \\
l_s \\
\mathcal{L}_{-}
\end{array}\right) \,,
\end{equation}

\begin{figure*}[!t]
    \centering
    \begin{subfigure}{0.475\textwidth}
        \centering
\includegraphics[width=\textwidth]{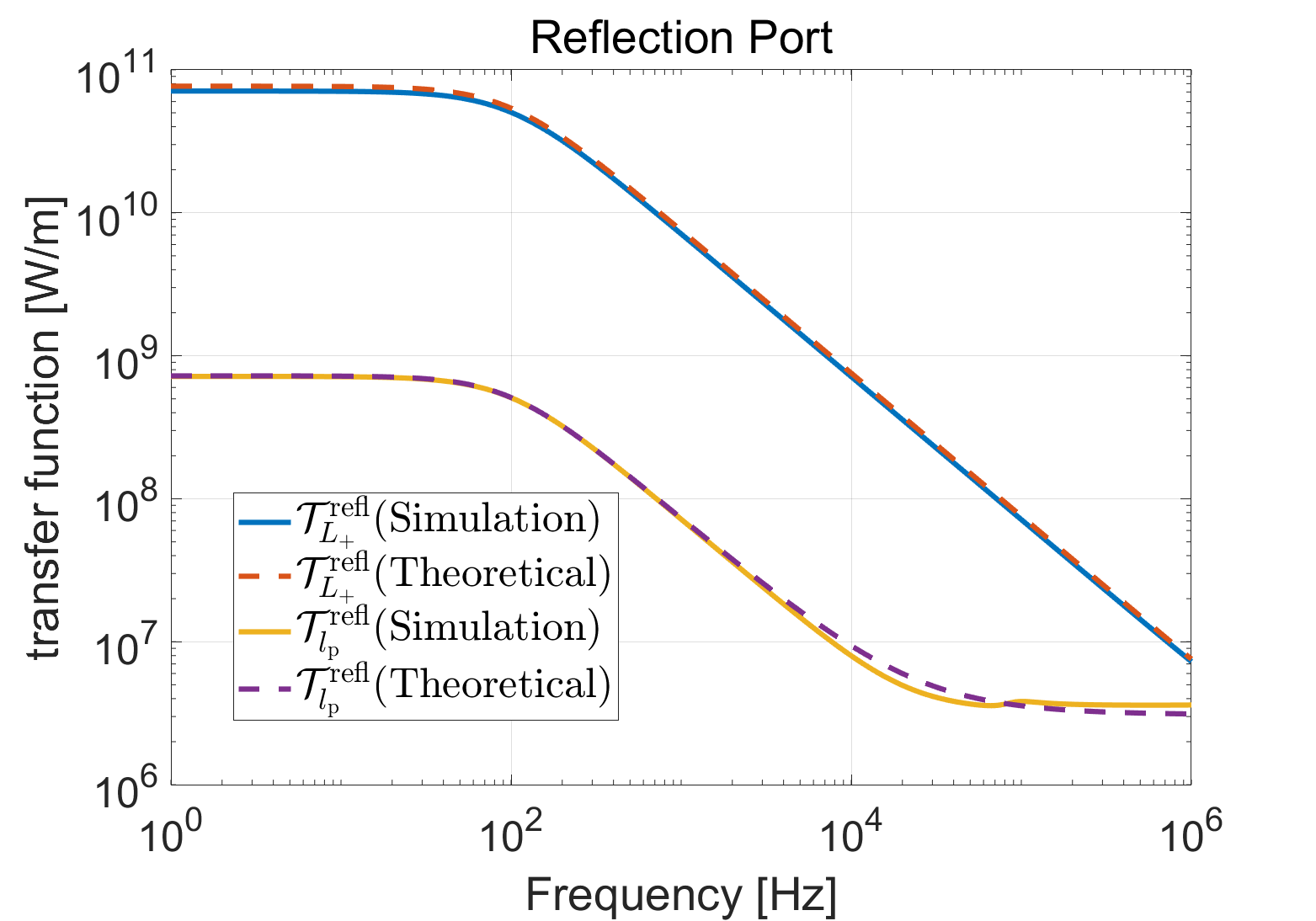}     
            \label{fig:tf1}
        \end{subfigure}
        \hfill
        \begin{subfigure}{0.475\textwidth}  
            \centering \includegraphics[width=\textwidth]{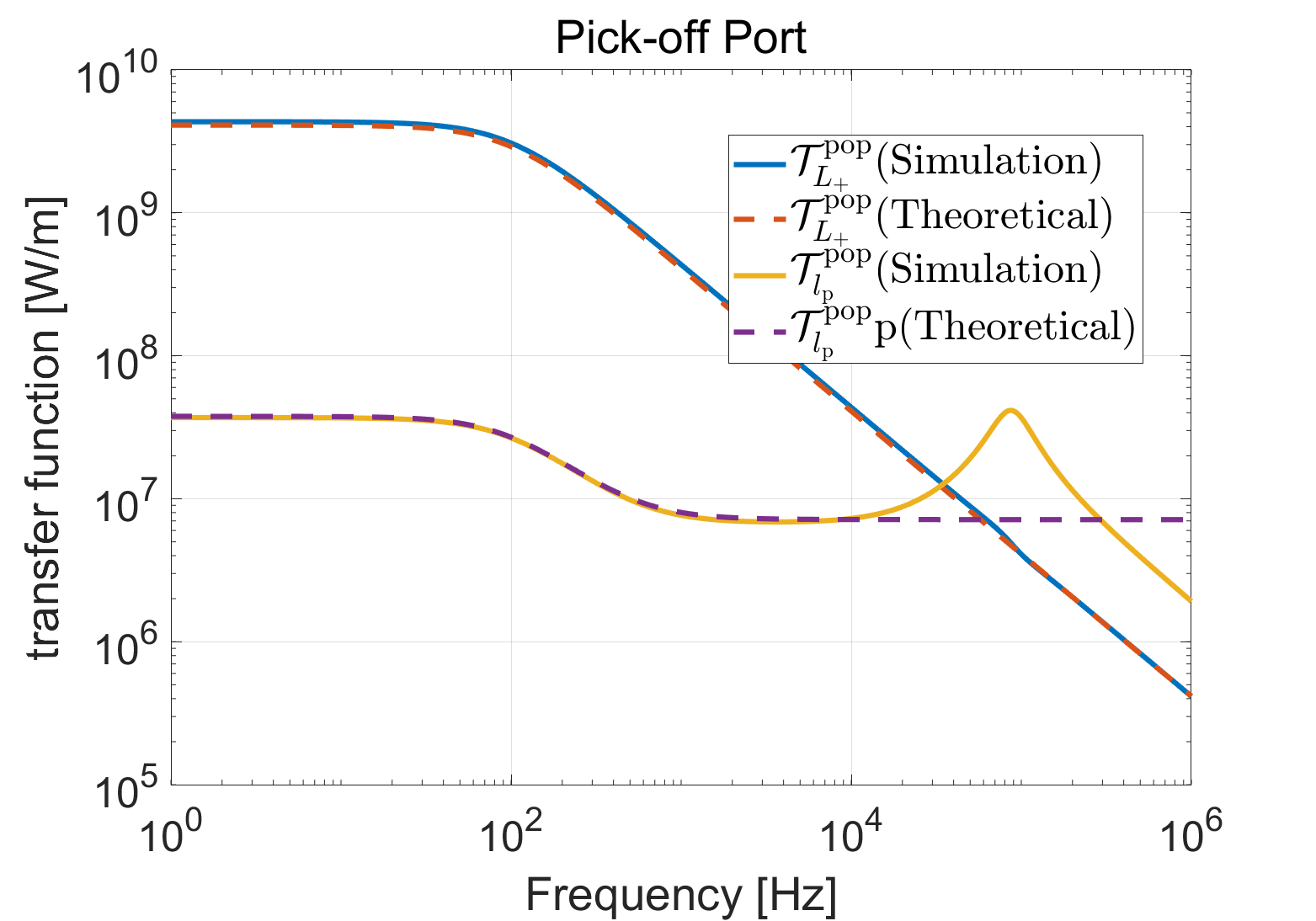}  
            \label{fig:tf2}
        \end{subfigure}
        \vskip\baselineskip
        \begin{subfigure}{0.475\textwidth}   
            \centering 
\includegraphics[width=\textwidth]{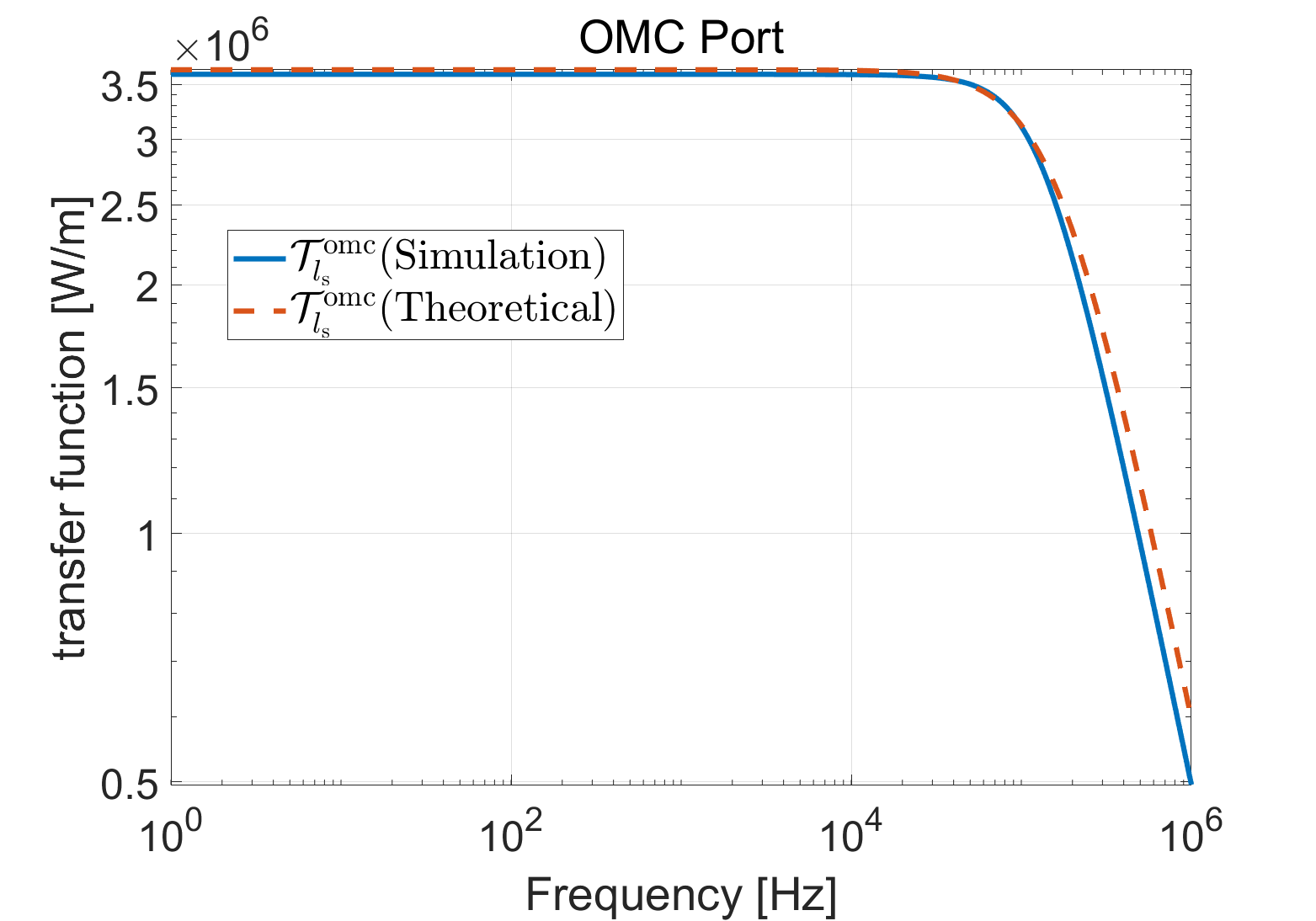}    
            \label{fig:tf3}
        \end{subfigure}
        \hfill
        \begin{subfigure}{0.475\textwidth}   
            \centering \includegraphics[width=\textwidth]{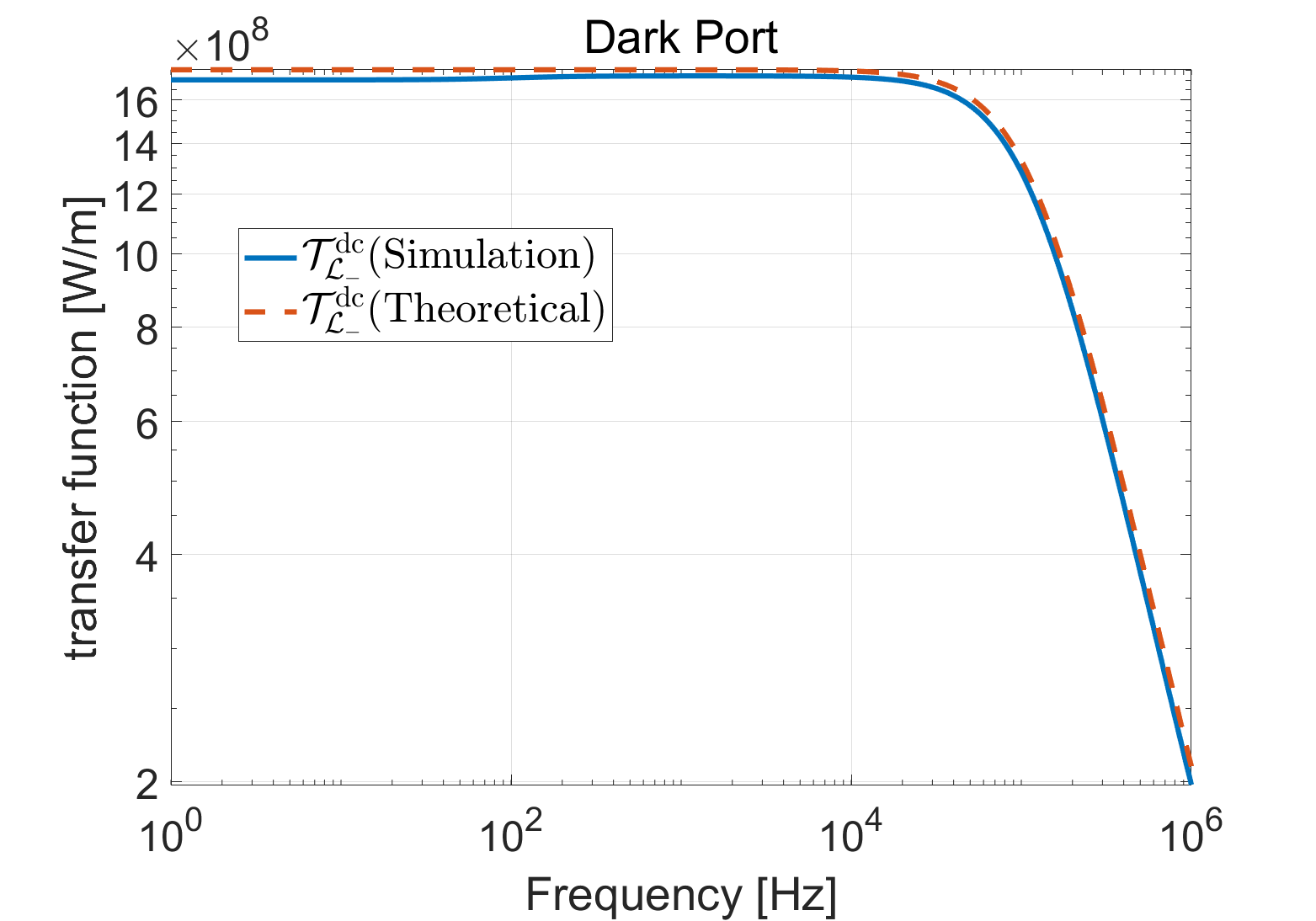}
            \label{fig:tf4}
        \end{subfigure}
        \caption[r]
        {\small Simulation results of non-zero transfer functions at different ports: Reflection port (top left); (b) Pick-off port (top right); (c) OMC port (bottom left); (d) Dark port (bottom right). They are compared with the analytical results summarized in Table\,\ref{tab:TFs}. The simulation is performed using Optickle \,\cite{Optickle}.} 
        \label{fig:TFs}
\end{figure*}

where $\mathcal{T}$ denotes the {length–to–power} transfer function, which quantifies how efficiently a displacement signal is converted into optical power at a given readout port. This function serves as a key figure of merit for assessing the effectiveness of a control strategy.  In practice, achieving transfer-function magnitudes of $|\mathcal{T}| \gtrsim 10^{6}\,\mathrm{W\,m^{-1}}$ for all relevant degrees of freedom is widely considered a necessary condition for the experimental viability of multi–degree-of-freedom locking schemes.

We derive the analytical expressions for the transfer functions based on the designed parameters; the results are summarized in Table\,\ref{tab:TFs}. These analytical predictions are further validated through numerical simulations, as illustrated in Fig.\,\ref{fig:TFs}.
Importantly, the achieved optical gain is sufficient to enable stable system locking.
While the high-frequency response (above 10\,kHz) exhibits slight deviations from the analytical predictions of Ref.\,\cite{Guo2023}—primarily due to the finite lengths of the recycling cavities—excellent agreement is maintained within the control bandwidth ($\leq 10$ kHz). This confirms that the experimental setup faithfully reproduces the control scheme envisioned for the full-scale interferometer, thereby establishing its effectiveness as a viable pathfinder. 

%% Section IV key elements
\section{Key components}
\label{secIV}

In this section, we provide a brief overview of the key components of the prototype, including the vacuum system, the seismic isolation system, and the digital control infrastructure.

 \begin{figure}
    \centering
   \includegraphics[width=0.85\columnwidth]{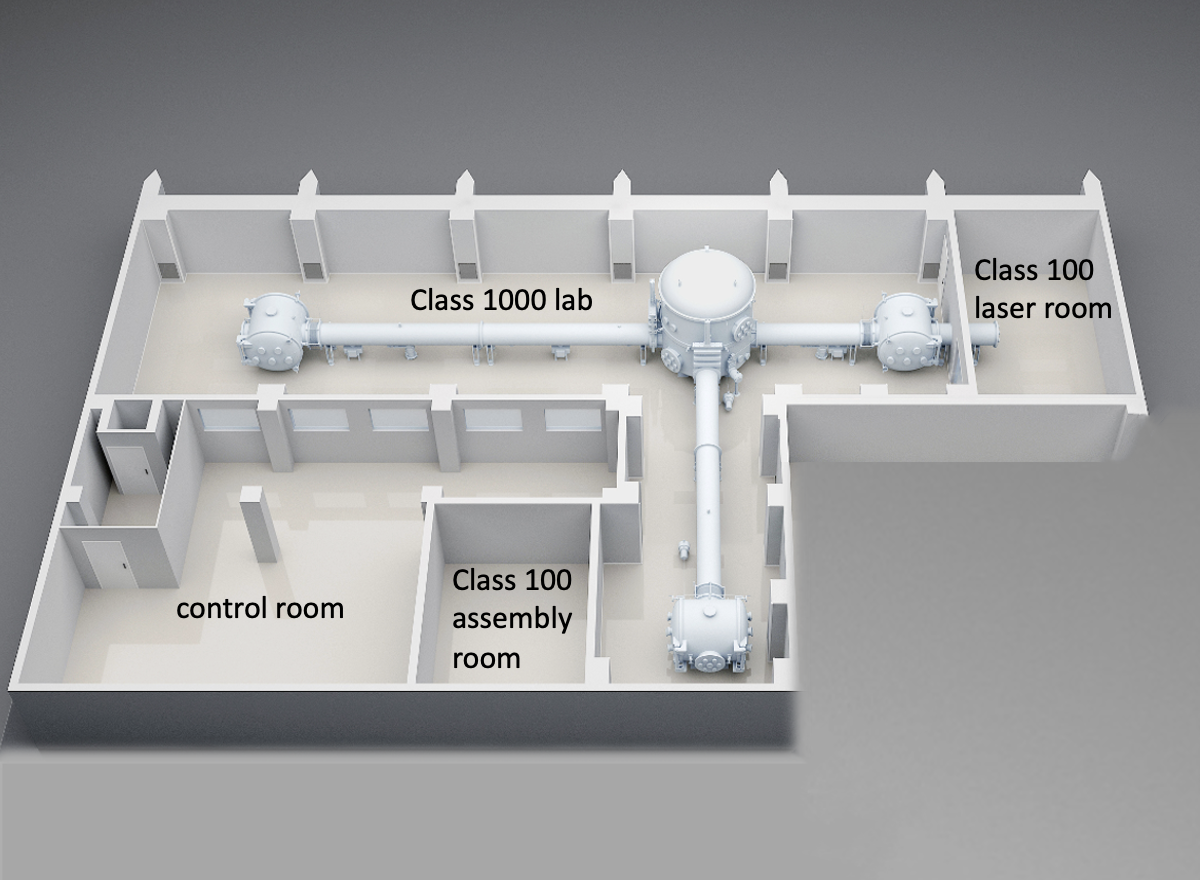}
    \caption{Layout of the BNU-12m prototype lab, which consists of a 260\,m$^2$ T-shape lab space and a 90\,m$^2$ control room at the southwest corner (up is North in this figure). The main lab has a Class 1000 cleanliness level and two separated rooms with higher cleanliness standards.}
    \label{fig:layout}
\end{figure}

\begin{figure}[htbp]
    \centering
    \begin{subfigure}{0.475\textwidth}
        \centering      \includegraphics[width=\textwidth, angle =0]{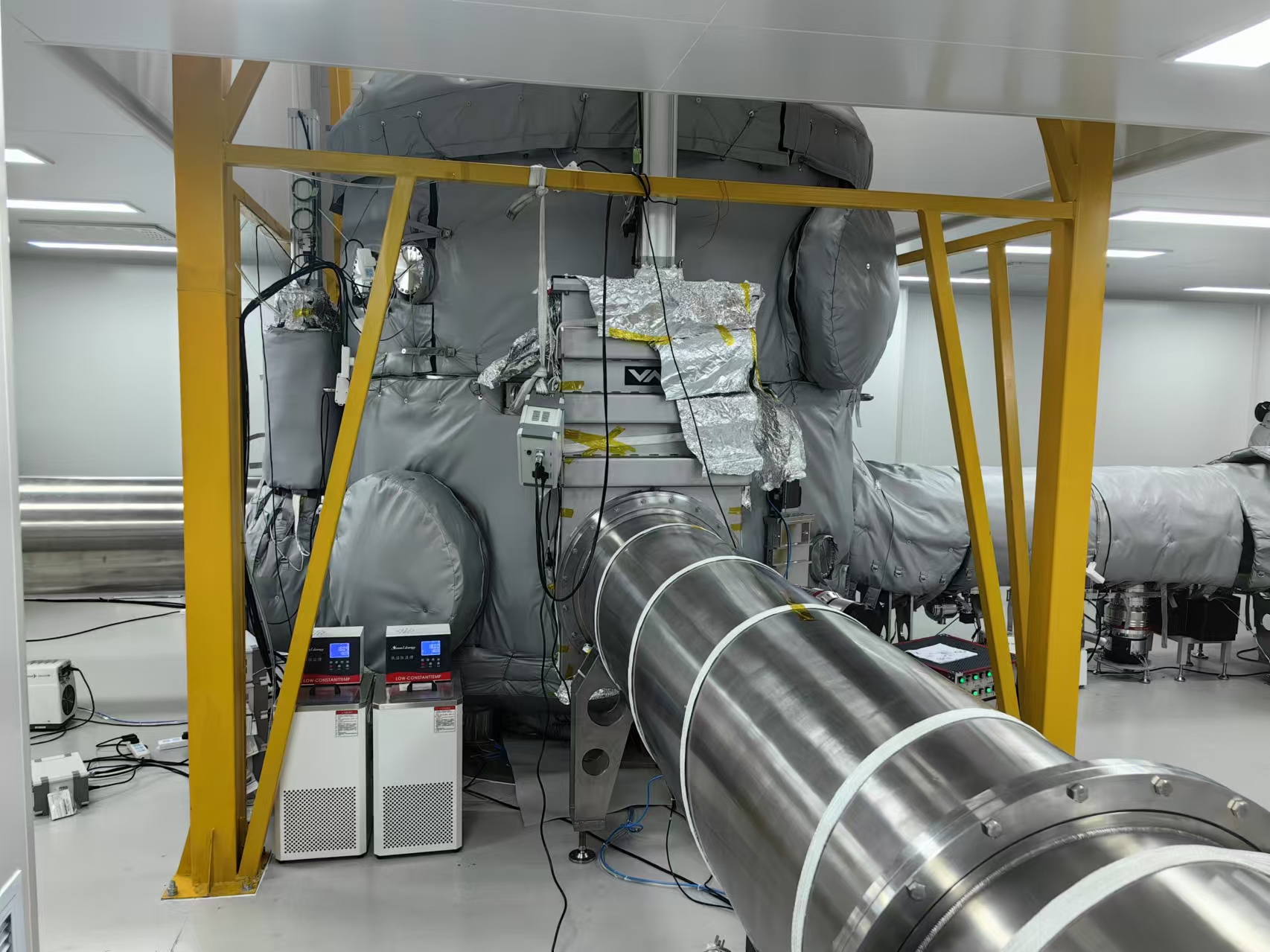}
            \caption[r]%
            {{\small Central tank featured in the lab, which has a lit that can be lifted up by the crane. It is 3.5 meters tall and 2.5 meters in diameter.}}    
            \label{fig:tank1}
        \end{subfigure}
        \hfill
        \begin{subfigure}{0.475\textwidth}  
            \centering         \includegraphics[width=\textwidth, angle =0]{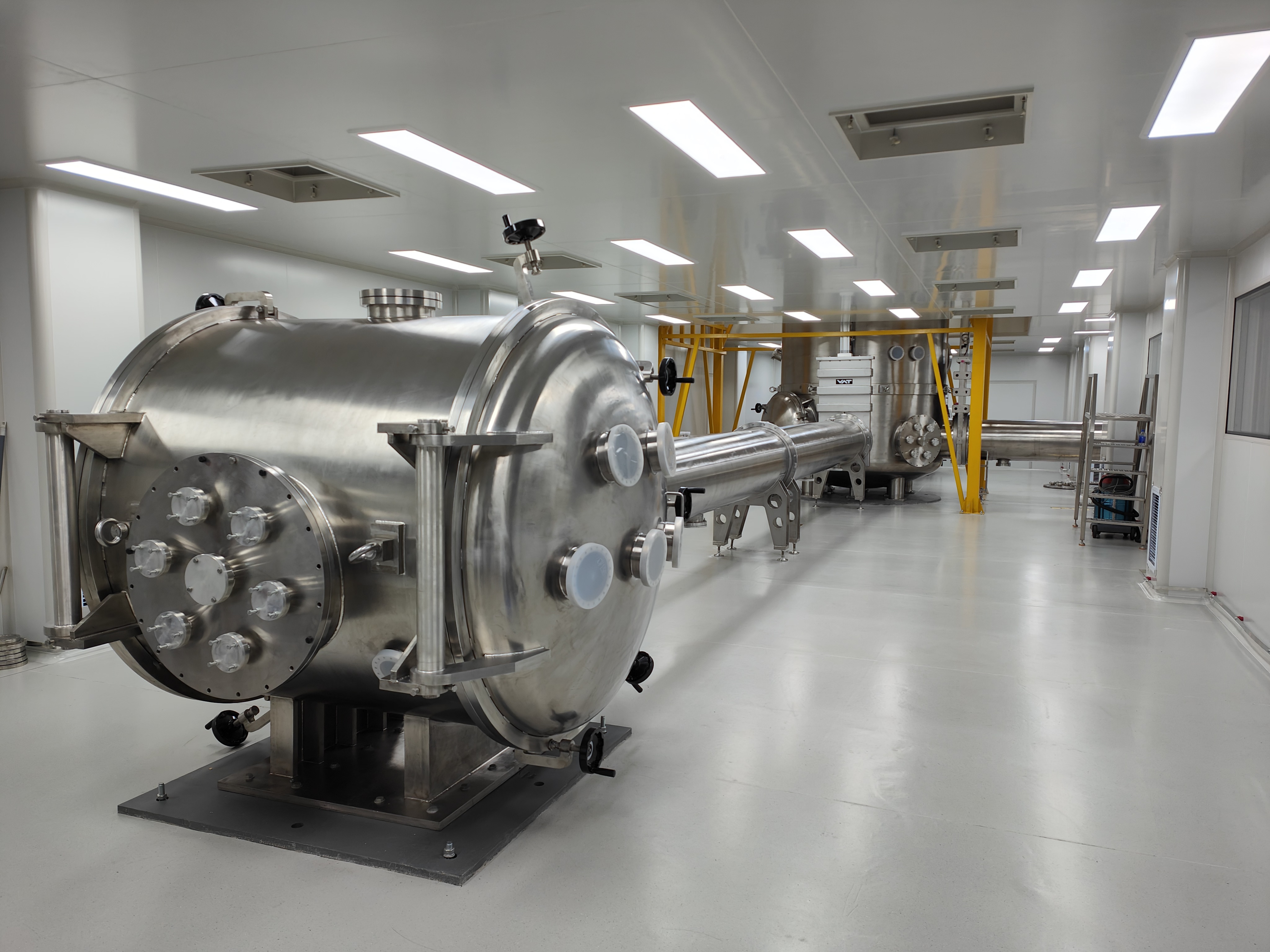}
            \caption[r]%
            {{\small Image of one of the three side tanks. All can be opened through both sides and have been fixed to their designated positions. }}    
            \label{fig:tank2}
        \end{subfigure}
        \caption[right]{The central and west tanks of the BNU 12m GW prototype facility.}
        \label{fig:vacuum}
\end{figure} 

\subsection{Vacuum system}

The vacuum system of the BNU prototype is housed in a $260\,\rm m^2$ T-shaped laboratory space, as illustrated in Fig.\,\ref{fig:layout}. It consists of a central tank measuring 2.5\,m in diameter and 3.5\,m in height, accompanied by three auxiliary side tanks. All tanks are accessible from the side, facilitating maintenance and alignment. The tanks are interconnected by vacuum tubes, each with a diameter of 60\,cm. The total volume of the vacuum system is approximately $30\,\rm m^3$. The system can be segmented into three distinct sections using two large gate valves positioned on the west and south sides of the central tank. The entire assembly has been securely installed in the laboratory, as shown in Fig.\,\ref{fig:vacuum}.

A three-stage vacuum pumping procedure is employed. The first stage uses four dry roots pumps, each with a pumping speed of $37\rm m^3/h$. The second stage employs four turbomolecular pumps, each with a pumping speed of $1250\,\rm L/s$, which significantly reduces the pressure beyond what dry pumps can achieve. The final stage utilizes ion pumps to maintain the ultra-high vacuum required for sensitive GW detection. The BNU prototype has demonstrated the ability to maintain a vacuum pressure of 
$5\times10^{-5}$
 Pa using six ion pumps, each with a pumping speed of $400\,\rm L/s$.

\subsection{Seismic isolation system}\label{sec:sus}

The seismic isolation system is critical for mitigating the impact of ground vibrations, which can hinder the lock acquisition of both the high-finesse L-shaped cavity and the coupled-cavity system. The linear operating range of the cavity is approximately 1\,nm, determined by the laser wavelength divided by the cavity finesse. Given that the feedback control bandwidth is limited to the kHz range—constrained by the mechanical response of the suspension system—the root mean square (RMS) velocity of the mirrors must remain below approximately 1000\,nm/s.
\begin{figure*}[t!]
    \centering
    \vspace{0.9cm}
    \begin{subfigure}{0.475\textwidth}
        \centering
        \includegraphics[width=\textwidth]{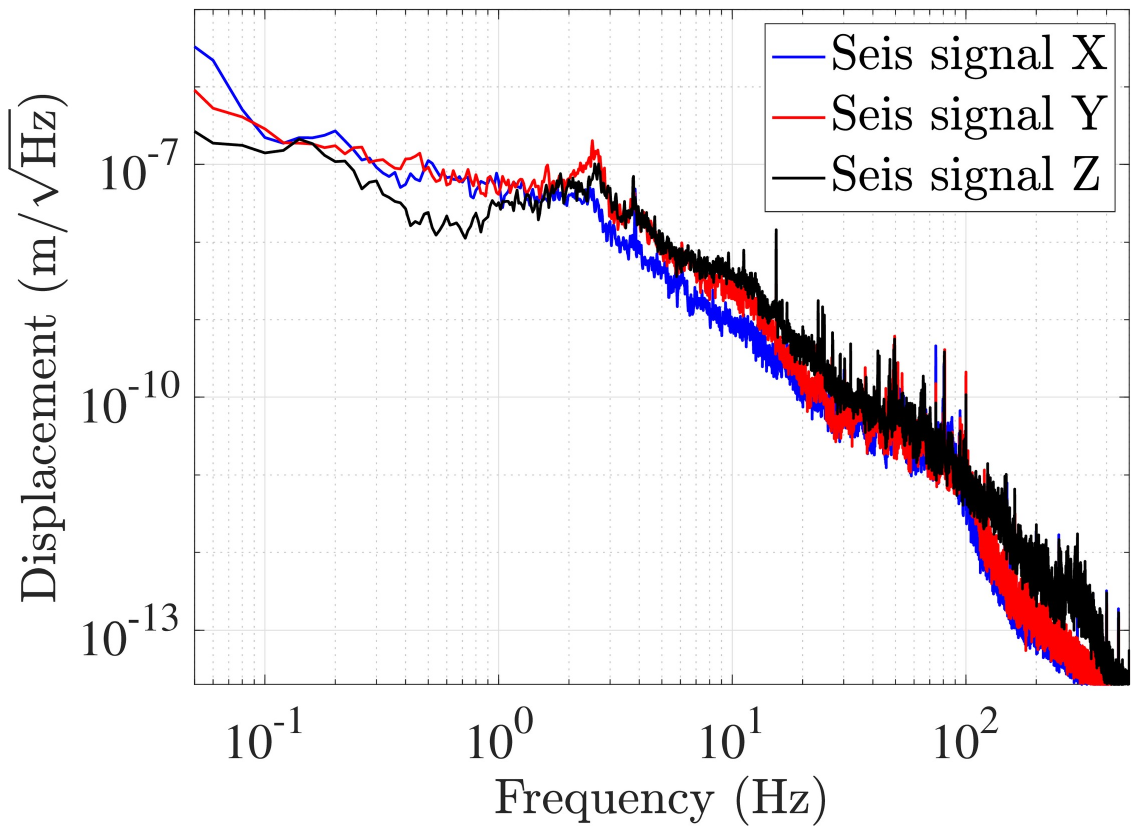}
            \caption[r]%
            {{\small Measured ground vibration noise in the laboratory. The seismic noise spectra along the  X, Y, and Z directions are shown.}}    
            \label{fig:vib1}
        \end{subfigure}
        \hfill
        \begin{subfigure}{0.475\textwidth}  
            \centering 
            \includegraphics[width=\textwidth]{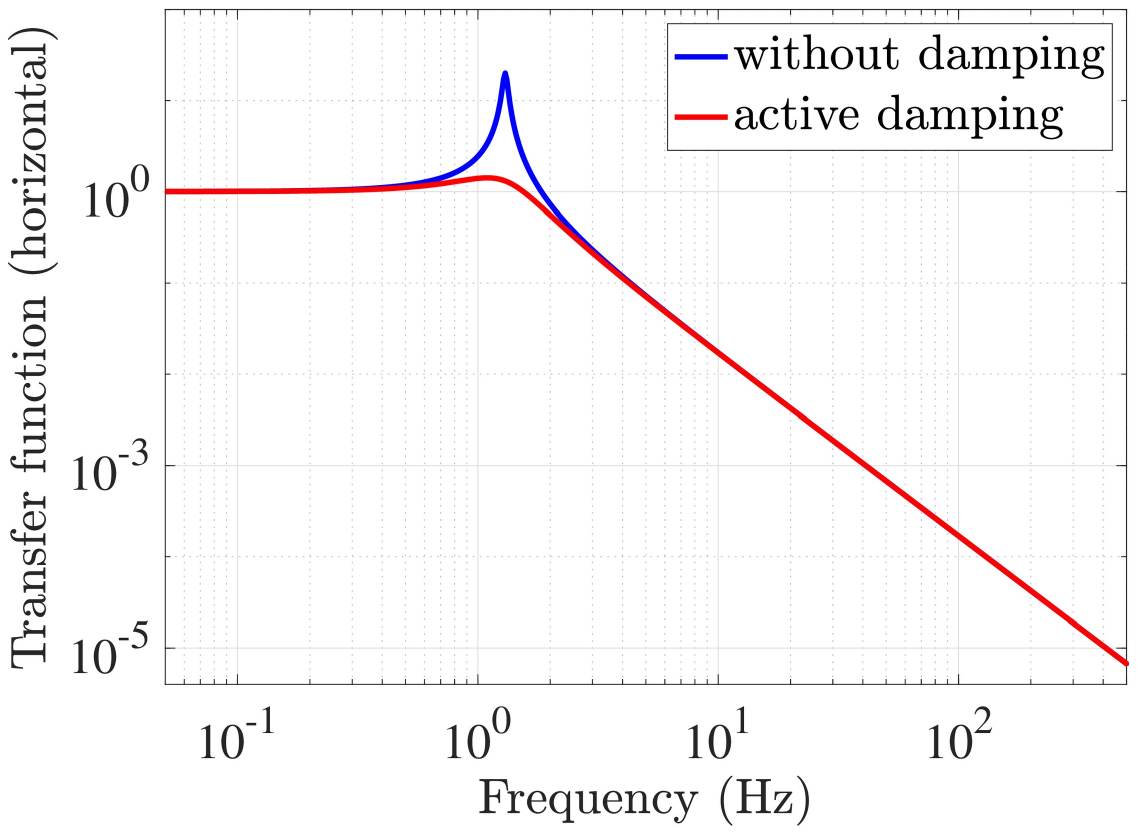}
            \caption[r]%
            {{\small Transfer function of the tip-tilt suspension (blue line), compared with the transfer function under active damping control.}}    
            \label{fig:vib2}
        \end{subfigure}
        \vskip\baselineskip
        \begin{subfigure}{0.475\textwidth}   
            \centering 
            \includegraphics[width=\textwidth]{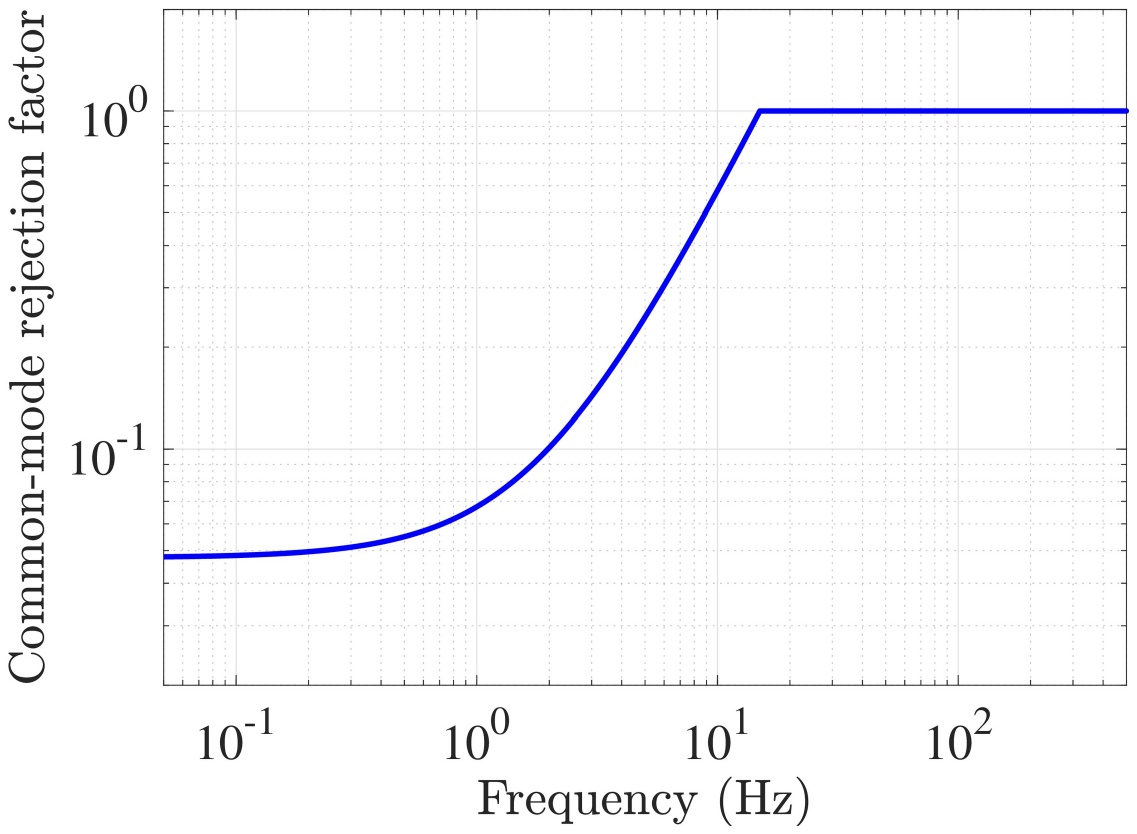}
            \caption[r]%
            {{\small Common-mode noise suppression effect, effectively filtering out seismic disturbances  below 15\,Hz.}}    
            \label{fig:vib3}
        \end{subfigure}
        \hfill
        \begin{subfigure}{0.475\textwidth}   
            \centering 
            \includegraphics[width=\textwidth]{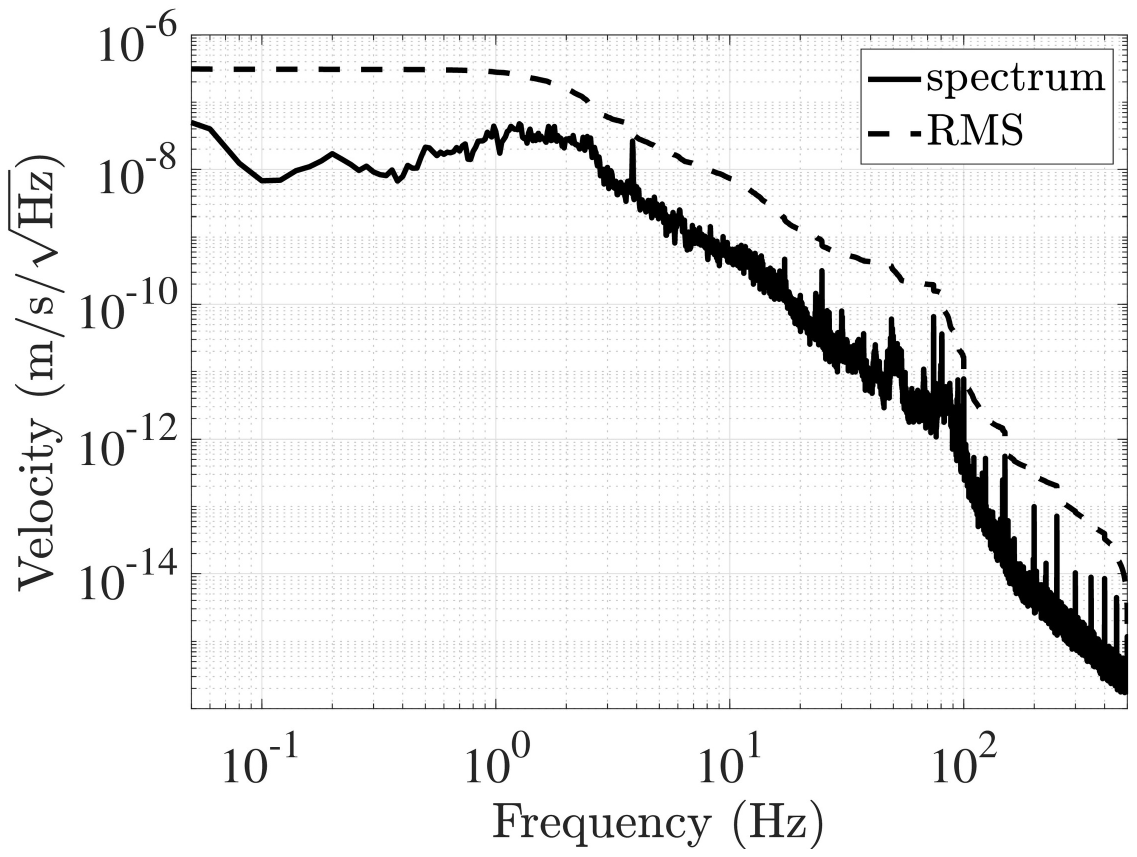}
            \caption[r]%
            {{\small Estimated velocity spectrum and root mean square (RMS) value of the mirrors in the optical resonant cavity.}}
            \label{fig:vib4}
        \end{subfigure}
        \caption[r]
        {\small Impact of ground vibration on the mirror velocity of the optical resonant cavity.} 
        \label{fig:vib}
\end{figure*}

The prototype employs single-stage LIGO-type Tip-Tilt suspensions for suspending the core optics\,\cite{tiptilt}, which satisfy the requirements for achieving stable cavity locking. The ground vibration level in the laboratory has been measured using a T120 seismometer, with the resulting noise spectrum shown in Fig.\,\ref{fig:vib1}. Using the transfer function of the tip-tilt suspension (Fig.\,\ref{fig:vib2}) and the estimated common-mode suppression factor (Fig.\,\ref{fig:vib3}), the resulting relative RMS velocity of the suspended optics is estimated to be approximately $200\,\rm nm/s$ (see Fig.\,\ref{fig:vib4}). 
This value is close to the ideal. However, when combined with a power recycling cavity, it poses challenges for coupled-cavity locking and underscores the need to explore and develop more sophisticated seismic isolation strategies for future upgrades.

\subsection{Digital control system}

Sophisticated electronics and control systems are essential for maintaining the alignment and locking of the entire interferometer, as well as for reading out and interpreting the data. The BNU prototype will adopt the LIGO Control and Data System (CDS)\,\cite{ligocds} to manage the entire interferometer, leveraging the proven capabilities of this advanced system to ensure precision and reliability. By utilizing a system based on the LIGO CDS, the BNU prototype aligns itself with the global GW research community. This alignment facilitates seamless data sharing and collaboration with other observatories and research teams. Moreover, it also provides structured training opportunities for researchers and students involved in the BNU project. 

%% Section V MHz detection

\section{MHz GW detection}\label{secV}
\begin{figure}[t!]
\centering
    \begin{minipage}[t]{0.49\textwidth}
        \centering
        \includegraphics[width=\textwidth]{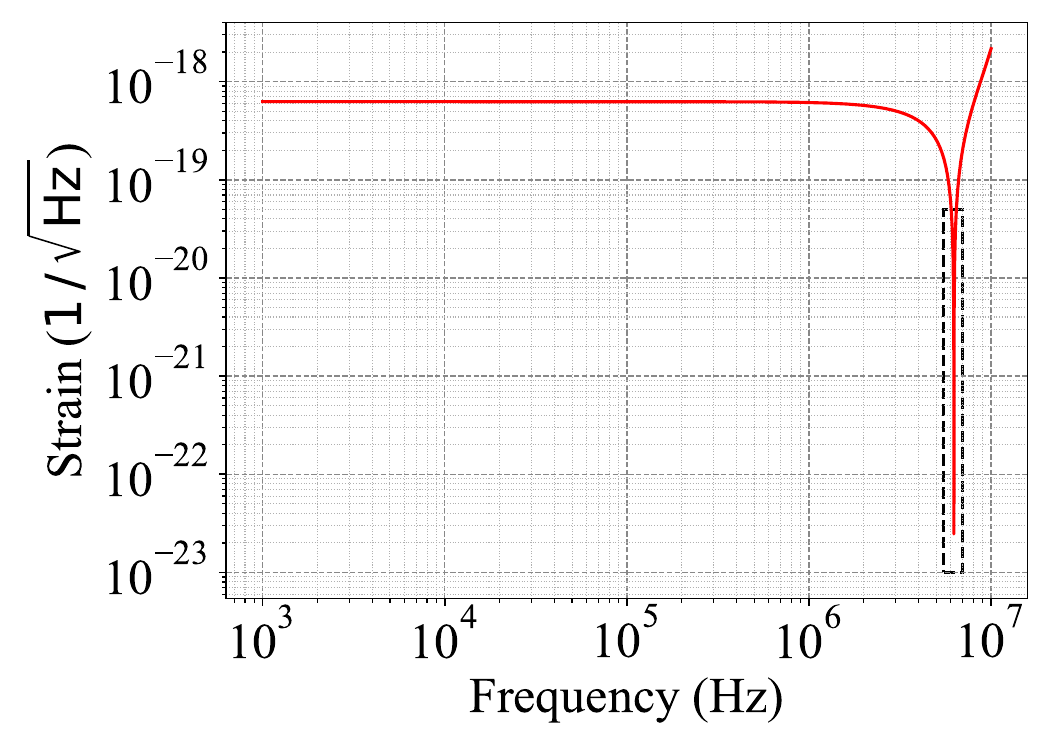}

    \end{minipage}
    \begin{minipage}[t]{0.49\textwidth}
        \centering
        \includegraphics[width=\textwidth]{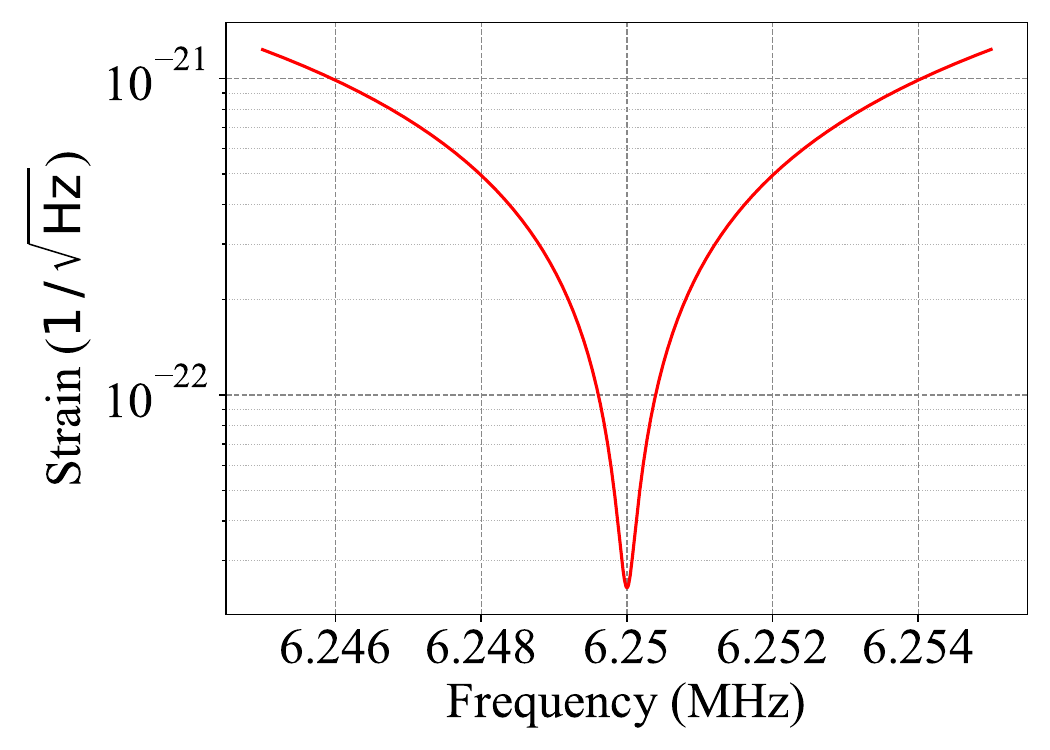}
    \end{minipage}

\caption[r]
        {\small Broadband (left) and zoomed-in (right) sensitivity of the BNU prototype. } 
\label{fig:noisebudget}
\end{figure}
Apart from its primary scientific goal of testing the sensing and control scheme for the new configuration, the prototype is also capable of achieving reasonable sensitivity in the MHz regime. This is enabled by its ten-meter-scale optical cavity, which supports an optical resonance near the MHz band.
As shown in Fig.~\ref{fig:noisebudget}, the prototype exhibits a peak sensitivity of approximately $3.0\times 10^{-23}/\sqrt{\rm Hz}$ at $6.25\rm MHz$, assuming an intra-cavity power of 300 kW and 5 dB of phase squeezing.

We explore the prototype’s sensitivity in detecting GWs from two potential exotic sources in the MHz regime (See Ref.\,\cite{Aggarwal:2020olq} for a comprehensive review of MHz-band sources and detection techniques).  The first example is a binary system composed of equal-mass primordial black holes. The corresponding signal-to-noise ratio (SNR) is given by  
\begin{equation}
\begin{aligned}
{\rm SNR} &=2 \sqrt{\int^{f_{\rm max}}_{f_{\rm min}}df \frac{|h(f)|^2}{S_{hh}(f)}}\\
&\approx 2.1 \times\left(\frac{ M}{10^{-3}{M_{\odot}}}\right)^{5/6}\times \left(\frac{d}{1\,\rm pc}\right)^{-1}\,.
\end{aligned}
\end{equation}
Here $(f_{\rm min},f_{\rm max})$ is symmetrically centered around  $6.25$ MHz with a bandwidth of 10 kHz; the waveform  $h(f)$ is computed using the IMRPhenomB model following Ref.\,\cite{PhysRevLett.106.241101}; $S_{hh}$ denotes the strain noise spectral density, whose amplitude is shown in Fig.\,\ref{fig:noisebudget}. 
This result suggests that the prototype would only be able to detect such binary systems within distances on the order of a parsec. Therefore, it is unlikely to observe primordial black hole binaries, as current GW detectors
---being sensitive at lower frequencies where $h(f) \propto f^{-7/6}$---would have already detected these signals with much higher SNRs.

The second potential source is geontropic fluctuation—an effect potentially arising from unknown quantum gravity phenomena—which manifests as a stochastic GW-like background\,\cite{PhysRevD.107.024002, photon_counting}.  The 
SNR for detecting such a stochastic signal is 
\begin{align}
{\rm SNR} &= \sqrt{\int^{f_{\rm max}}_{f_{\rm min}}T_{\rm int}\cdot df \frac{S_{h}^2(f)}{S_n^2(f)}} \\
&\approx  32.1\times \left(\frac{\alpha}{0.1}\right)\times\left(\frac{T_{\rm int}}{1 \rm s}\right)^{\frac{1}{2}}\,,
\end{align}
where $T_{\rm int}$ is the integration time and 
$\alpha$ is the normalized coupling coefficient.  
This result suggests that it may be possible to detect geontropic fluctuations if the coupling coefficient $\alpha$ of the order of unity. 
 
\newpage
\section{Conclusion}
\label{sect:conclusion}
The BNU 12-meter prototype represents a dedicated effort toward realizing the next generation of gravitational-wave detectors with enhanced sensitivity in the kilohertz regime. By implementing an L-shaped optical resonator as the core sensing element, the prototype seeks to overcome fundamental limitations of traditional Michelson-based interferometers—particularly those arising from quantum noise and optical loss in the high-frequency band. The experimental configuration leverages a power-recycled, signal-extracted coupled cavity design, with suspended optics enclosed in a custom-built vacuum system and supported by a seismic isolation strategy sufficient to enable stable locking at high finesse.

Through a staged implementation plan, the prototype focuses first on validating the locking and control of a high-finesse suspended L-shaped cavity, followed by the integration of the full coupled cavity system including the power recycling and signal extraction mirrors. Preliminary simulations and measured environmental data indicate that the proposed control schemes and optical configurations are feasible within the current technical framework. The use of LIGO-style tip-tilt suspensions, coupled with active damping and common-mode noise suppression, ensures sufficiently low residual motion for stable operation. The prototype also adopts the LIGO Control and Data System, integrating it into the global GW detection infrastructure and enabling cross-collaboration and data sharing with the broader community.

Looking forward, as the BNU prototype progresses through its development stages, the prototype will yield valuable insights into optical design, control strategies, and noise mitigation techniques, all of which are essential for scaling to full-scale kHz detectors. 

\acknowledgements
The authors would like to thank Kenneth Strain, Denis Martynov, Jerome Degallaix and Conor Mow-Lowry for helpful discussions. This work is supported by the National Key Research and Development Program of China, Grant/Award Number: 2024YFC2208000 and 2023YFC2205800; National Natural Science Foundation of China under Grant No.12021003, No.12474481 and No.123B1015; Fundamental Research Funds for the Central Universities; JST ASPIRE Program, Japan Grant No. JPMJAP2320.

\section*{Appendix: key quantifies related to transfer functions}
\label{sec:appendixA}

In this appendix, we summarize the key quantities used in the analytical expressions for the optical transfer functions listed in Table\,\ref{tab:TFs}. 
These include the effective optical gains and reflectivities of the power recycling cavity (PRC) and signal extraction cavity (SEC). We label the auxiliary cavities with subscripts $\rm p$ and $\rm s$ corresponding to the PRC and SEC, respectively, and distinguish between carrier and sideband fields using superscripts $\rm c$ and $\rm s$. The optical gains take the form:
\begin{equation}
\begin{aligned}\label{eq9}
g_\textrm{p}^{\rm c}&=\frac{t_{\rm p}}{1+t_{\rm po}^2r_{\rm p}r_\textrm{L,\,c}^{(+)}}, 
\; g_\textrm{p}^{\rm s}=\frac{r_{\rm p}}{1-t_{\rm po}^2r_{\rm p}r_\textrm{L,\,s}^{(+)}},\;\\ 
g_\textrm{s}^{\rm c}&=\frac{t_{\rm s}}{1+r_{\rm s}r_\textrm{L,\,c}^{(-)}},\;\quad \,\,g_\textrm{s}^{\rm s}=\frac{t_{\rm s}}{1+r_{\rm s}r_\textrm{L,\,s}^{(-)}}\,, 
\end{aligned}
\end{equation}
where $r_{\rm p}$ and $r_{\rm s}$ are the reflectivity of PRM and SEM respectively, and $t_{\rm po}$ is the amplitude transmissivity of the pick-off mirror. The effective reflectivities of the system as seen from the bright port and dark port are given by:
\begin{align}\label{eq11}
r_\textrm{p}^{\rm c} &= r_{\rm p}+r_\textrm{L,c}^{(+)}t_pg_\textrm{p}^{\rm c} \;,r_\textrm{p}^{\rm s} = r_{\rm p}-r_\textrm{L,s}^{(+)}t_pg_\textrm{p}^{\rm s} \;,
r_\textrm{s}^{\rm s} = r_{\rm s}+r_\textrm{L,s}^{(-)}t_sg_\textrm{s}^{\rm s}\,.
\end{align}
In these expressions, the effective reflectivities for the carrier and sidebands are defined as:
\begin{equation}
\begin{aligned}
r_{\mathrm{L}, \rm c}^{(\pm)}=-\frac{t_{\mathrm{i}}^2 r_{\mathrm{e}}}{1 \mp r_{\mathrm{i}} r_{\mathrm{e}}} \pm r_{\mathrm{i}}\,; \,
r_{\mathrm{L}, \rm s }^{(\pm)}=-\frac{t_i^2 r_{\mathrm{e}} e^{i \phi_m}}{1 \mp r_{\mathrm{i}} r_{\mathrm{e}} \mathrm{e}^{i \phi_m}} \pm r_{\mathrm{i}}\,,
\end{aligned}
\end{equation}
where $r_{\rm i}$ and $r_{\rm e}$ stands for the reflectivity of ITM and ETMs. The subscripts $c$ and $s$ denote carrier and sidebands, while $+$ and $-$ denote looking from the bright port and the dark port, respectively. The sideband modulation phase is denoted by $\phi_m \equiv 2\omega_{m_i} L_+/c$, where $i=1,2$ labels the bright-port sideband and the auxiliary laser.

% Within the control bandwidth, the non-zero transfer functions at OMC port and dark port are basically flat.  share an identical single–pole response whose corner frequency,
% $f_{\mathrm{cc}}$, coincides with the coupled–cavity bandwidth set by the simultaneous resonance of the arm cavities and the power-recycling cavity.  
% Meanwhile, the $l_{p}$ and $L_{+}$ transfer functions obtained from the reflection port and the pick-off port exhibit a more intricate, fractional-linear frequency dependence; moreover, their poles occur at distinct frequencies.  This separation of poles ensures that the total power readouts at the reflection and pick-off ports remain linearly independent.
\bibliography{sn-bibliography}

\begin{thebibliography}{10}

\bibitem{2016PhRvL.116f1102A}
B.~P. {Abbott}, R.~{Abbott}, et~al.
\newblock {Observation of Gravitational Waves from a Binary Black Hole Merger}.
\newblock {\em Phys. Rev. Lett.}, 116(6):061102, February 2016.

\bibitem{PhysRevLett.119.161101}
B.~P. Abbott, R.~Abbott, et~al.
\newblock Gw170817: Observation of gravitational waves from a binary neutron star inspiral.
\newblock {\em Phys. Rev. Lett.}, 119:161101, Oct 2017.

\bibitem{AdLigo2015}
J~Aasi, B~P Abbott, et~al.
\newblock Advanced ligo.
\newblock {\em Classical and Quantum Gravity}, 32:074001, 4 2015.

\bibitem{Abbott_2017}
B.~P. Abbott, R.~Abbott, et~al.
\newblock Multi-messenger observations of a binary neutron star merger.
\newblock {\em The Astrophysical Journal Letters}, 848(2):L12, oct 2017.

\bibitem{Abbott_HC}
B.~P. Abbott, R.~Abbott, et~al.
\newblock A gravitational-wave standard siren measurement of the hubble constant.
\newblock {\em Nature}, 551(7678):85--88, 2017.

\bibitem{Martynov19}
Denis Martynov, Haixing Miao, et~al.
\newblock Exploring the sensitivity of gravitational wave detectors to neutron star physics.
\newblock {\em Phys. Rev. D}, 99:102004, May 2019.

\bibitem{Bailes21}
M.~Bailes, B.~K. Berger, et~al.
\newblock Gravitational-wave physics and astronomy in the 2020s and 2030s.
\newblock {\em Nature Reviews Physics}, 3(5):344--366, 2021.

\bibitem{CMB}
S.~Khlebnikov and I.~Tkachev.
\newblock Relic gravitational waves produced after preheating.
\newblock {\em Phys. Rev. D}, 56:653--660, Jul 1997.

\bibitem{AmaroSeoane2023}
Pau Amaro-Seoane, Jeff Andrews, et~al.
\newblock Astrophysics with the laser interferometer space antenna.
\newblock {\em Living Reviews in Relativity}, 26(1):2, 2023.

\bibitem{tianqin}
Jun {Luo}, Li-Sheng {Chen}, et~al.
\newblock {TianQin: a space-borne gravitational wave detector}.
\newblock {\em Classical and Quantum Gravity}, 33(3):035010, February 2016.

\bibitem{Taiji}
Ziren {Luo}, ZongKuan {Guo}, et~al.
\newblock {A brief analysis to Taiji: Science and technology}.
\newblock {\em Results in Physics}, 16:102918, March 2020.

\bibitem{Li2023}
Junlang Li, Fangfei Liu, et~al.
\newblock Detecting gravitational wave with an interferometric seismometer array on lunar nearside.
\newblock {\em Science China Physics, Mechanics \& Astronomy}, 66(10):109513, 2023.

\bibitem{zhang2023}
Teng Zhang, Huan Yang, et~al.
\newblock Gravitational-wave detector for postmerger neutron stars: Beyond the quantum loss limit of the fabry-perot-michelson interferometer.
\newblock {\em Phys. Rev. X}, 13:021019, May 2023.

\bibitem{Caltech40m}
LIGO~Scientific Collabration.
\newblock Ligo caltech 40 meter prototype, https://labcit.ligo.caltech.edu/~ajw/40m/40m-upgrade.html.
\newblock 2007.

\bibitem{Goßler_2010}
S~Goßler, A~Bertolini, et~al.
\newblock The aei 10 m prototype interferometer.
\newblock {\em Classical and Quantum Gravity}, 27(8):084023, apr 2010.

\bibitem{Zhao_2006}
C~Zhao, D~G Blair, et~al.
\newblock Gingin high optical power test facility.
\newblock {\em Journal of Physics: Conference Series}, 32(1):368, mar 2006.

\bibitem{Glasgow10}
D.~I. Robertson, E.~Morrison, et~al.
\newblock The glasgow 10m prototype laser interferometric gravitational wave detector.
\newblock {\em Review of Scientific Instruments}, 66(9):4447--4452, 09 1995.

\bibitem{Aggarwal:2020olq}
Nancy Aggarwal et~al.
\newblock {Challenges and opportunities of gravitational-wave searches at MHz to GHz frequencies}.
\newblock {\em Living Rev. Rel.}, 24(1):4, 2021.

\bibitem{Guo2023}
Xinyao Guo, Teng Zhang, et~al.
\newblock Sensing and control scheme for the inteferometer configuration with an l-shaped resonator.
\newblock {\em Classical and Quantum Gravity}, 40(23):235005, oct 2023.

\bibitem{tiptilt}
Bram J.~J. Slagmolen, Adam~J. Mullavey, et~al.
\newblock {Tip-tilt mirror suspension: Beam steering for advanced laser interferometer gravitational wave observatory sensing and control signals}.
\newblock {\em Review of Scientific Instruments}, 82(12):125108, 12 2011.

\bibitem{Optickle}
Evans M.
\newblock {\em Technical Report No. T070260-v1 LIGO lab}, 2007.
\newblock \url{https://dcc.ligo.org/LIGO-T070260/public/}.

\bibitem{ligocds}
R.~Bork, R.~Abbott, et~al.
\newblock An overview of the ligo control and data acquisition system, arxiv:physics/0111077.
\newblock 2001.

\bibitem{PhysRevLett.106.241101}
P.~Ajith, M.~Hannam, et~al.
\newblock Inspiral-merger-ringdown waveforms for black-hole binaries with nonprecessing spins.
\newblock {\em Phys. Rev. Lett.}, 106:241101, Jun 2011.

\bibitem{PhysRevD.107.024002}
Dongjun Li, Vincent S.~H. Lee, et~al.
\newblock Interferometer response to geontropic fluctuations.
\newblock {\em Phys. Rev. D}, 107:024002, Jan 2023.

\bibitem{photon_counting}
Lee McCuller.
\newblock Single-photon signal sideband detection for high-power michelson interferometers, 2022.

\end{thebibliography}
\bibliographystyle{unsrt85}

\end{document}